\documentclass[prb,twocolumn,showpacs]{revtex4}
\input epsf
\epsfverbosetrue

\usepackage{graphics}
\usepackage{amsmath}
\usepackage{bm}
\newcommand{\Journal}[4]{#1 \textbf{#2}, #3 (#4)}
\newcommand{\PRev}{Phys. Rev.}
\newcommand{\JPSJ}{J. Phys. Soc. Jpn.}
\newcommand{\be}{\begin{eqnarray}}
\newcommand{\ee}{\end{eqnarray}}
\newcommand{\nn}{\nonumber}

\newcommand{\pr}{PrOs$_4$Sb$_{12}$}

\begin{document}
\title{Impurity induced density of states and residual transport in nonunitary 
superconductors}
\author{T. R. Abu Alrub and S. H. Curnoe}
\affiliation{Department of Physics and Physical Oceanography,
Memorial University of Newfoundland, St. John's, NL, A1B 3X7, Canada}

\begin{abstract}
We obtain general expressions for the residual density of states, electrical
conductivity and thermal conductivity for non-unitary superconductors 
due to impurity scattering.
We apply the results to the so-called `B phase' of \pr, which we describe 
using a non-unitary gap function derived from symmetry considerations. 
The conductivity tensor has
inequivalent diagonal components due to off-axis nodal 
positions which may be detectable in experiments.
\end{abstract}

\pacs{74.20.-z, 71.27.+a, 71.10.-w}

\maketitle

\section{Introduction}
Non-unitary pairing in superfluids was first described
by Leggett,\cite{Leggett1975} but the A$_1$ phase of  $^3$He is the 
only well-established example of this, so far.
However, recently non-unitary
pairing was observed in the heavy fermion superconductor
\pr\ by Aoki {\em et al}.\cite{Aoki2003}
A physically significant consequence of
non-unitary pairing is a lifting of the degeneracy of the
superconducting energy gap, so that two different energy gap branches,
both of which are
anisotropic,
are observable.
Multi-gap behaviour has been observed in 
\pr\cite{Measson2004,Seyfarth2005,Seyfarth2006,Yogi2006,Sakakibara2007,Turel2007,MacLaughlin2007}
but so far this has mainly been attributed to multi-band superconductivity, and
gap splitting due to non-unitary pairing has received little consideration,
in spite of numerous citations of Aoki {\em et al.}'s results.

Superconductivity in \pr\
is believed 
to be unconventional.\cite{Maple2001,Bauer2002,Izawa2003,Aoki2003,Chia2003,Huxley2004,
Frederick2005,Nishiyama2005,Higemoto2007,Katayama2007} The paired electrons are in a spin triplet configuration,\cite{Higemoto2007} and the superconducting state has 
broken time reversal symmetry and is non-unitary.\cite{Aoki2003}
Low
temperature power law behaviour, indicative of the presence of nodes in the 
gap function, has been observed in thermodynamic and transport measurements,\cite{Bauer2002,Frederick2005,Chia2003,Izawa2003, Katayama2007}
but some  experiments have found the gap function to be 
nodeless.\cite{MacLaughlin2002,Kotegawa2003,Suderow2004,Seyfarth2006} 
Other experiments observed
two superconducting phases, possibly
with different symmetries,
\cite{Aoki2002,Aoki2003,Vollmer2003,Chia2003,Izawa2003,Tayama2003,Ho2003,
Oeschler2004,Grube2006} suggesting a multi-component 
superconducting order parameter.
These two phases are known as the ``A phase" and the ``B phase".  If it
exists, the A phase occupies only a small region of the phase diagram 
just below $H_{c2}(T)$.  Thus, most measurements, including
those cited above, have probed the B phase.

The three dimensional representation $T_u$ of the point group $T_h$ 
best describes superconductivity in \pr.\cite{Sergienko2004,Tayseer2} 
This representation 
yields several superconducting phases, of which four are 
accessible from the normal state by a second order phase transition. 
We have previously identified the states $D_2(C_2)\times {\cal K}$ and $D_2(E)$,
with order parameter components
$(0,0,|\eta_1|)$ and $(0, i|\eta_2|, |\eta_1|)$, as the 
A phase and 
B phase, respectively.\cite{Sergienko2004, Tayseer2}
Here, $D_2(C_2)$ is the symmetry group with elements
$\{E,C_2^x,
U(\pi)C_2^y,U(\pi)C_{2}^z\}$
while $D_2(E)= \{E,C_2^x{\cal K},
U_1(\pi)C_2^y{\cal K},U_1(\pi)C_2^z\}$.\cite{Sergienko2004}
The corresponding gap functions are unitary 
for the A phase, with two point nodes in the [00$\pm$1] directions, and nonunitary for the B phase,
with four nodes on unusual 
points on the Fermi surface, [0,$\pm\alpha$,$\pm\beta$]. 

Low temperature transport is an effective probe for the symmetry of 
the gap function.\cite{Lee1993,Hirschfeld1993,Graf1996,
Balatsky1994,Durst} 
Impurities induce and scatter quasiparticles
at the nodes and the conductance remains finite even in the limit
of zero frequency and temperature. 
Usually, two limiting cases of impurity scattering are considered, the
``Born limit" (weak scattering) and the ``unitary limit" (strong scattering).
The unitary limit is associated with non-magnetic substitutions of 
magnetic ions in heavy fermion superconductors.\cite{Pethick1986,Arfi1987,Hirschfeld1988,Schmitt-Rink1986}  
The self-energy due to isotropic impurity scattering is obtained from 
the T-matrix,\cite{Pethick1986,Hirschfeld1988}
$\Sigma(k,\omega) = (n_i/\pi N_n) T(k,k,\omega)$, where $n_i$ is the impurity
concentration, $N_n$ is the 
density of states in the normal state, and the T-matrix is the self-consistent solution
to $T(\omega) = V + V G_0(\omega) T(\omega)$, where
$V$ is the impurity potential, $G_0(\omega) = (1/\pi N_n) \sum_{\bm k} G({\bm k},\omega)$  
and $G({\bm k},\omega)$ is the electronic Green's
function in the superconducting state.
The self-energy is then $\Sigma(\omega) = (n_i/\pi N_n)G_0(\omega)/[c^2 - G_0^2(\omega)]$,
where $c$ is related to the phase shift, $c = \cot \delta_0$.
In unitary limit $c\rightarrow 0$, while $c \rightarrow \infty$ in the Born
limit.
The main result of this approach is a renormalisation of the
frequency $\omega \rightarrow \omega - i\Gamma(\omega)$
due to impurity scattering.
We will use this result to  find impurity induced residual density of
states and transport coefficients.

The outline of this paper is as follows: in Sec.\ $\rm{II}$ we define the gap function,
the mean field Green's functions and spectral functions.
In Secs.\ 
$\rm{III}$, $\rm{IV}$ and $\rm{V}$ we derive general expressions for 
the impurity induced quasiparticle 
density of states, the electrical conductivity and the thermal conductivity in
a nonunitary superconducting state.  In Sec.\ $\rm{VI}$ we apply
our results to the nonunitary B phase 
in \pr, and we summarise our results  in Sec.~$\rm{VII}$. 
\section{Mean Field Results}
In the following we state the main results of the mean field treatment
of an effective pairing Hamiltonian (see Ref.\ \onlinecite{Sigrist1991} for
details).

The gap function is a $2\times 2$ matrix in pseudospin space.  For
triplet pairing it can be parametrised in terms of an
odd pseudovectorial function ${\bm d}({\bm k})$ as
\begin{equation}
\label{2.2}
{\widetilde\Delta}_{\bm k}=i[{\widetilde{\bm\sigma}}\cdot{\bm d}_{\bm k}]
\widetilde\sigma_y = 
 \left( \begin{array}{cc}
-d_x({\bm k}) + id_y({\bm k}) & d_z({\bm k})\\
d_z({\bm k}) & d_x({\bm k}) + id_y({\bm k}) 
\end{array} \right).
\end{equation} 
When ${\widetilde \Delta}_{\bm k}{\widetilde \Delta}_{\bm k}^{\dagger}$
is proportional to the unit matrix, the pairing is said to be ``unitary".
Non-unitary pairing occurs only in the triplet channel and only when
${\bm q}_{\bm k} \equiv i{\bm d}_{\bm k}\times{\bm d}_{\bm k}^{*} \neq 0$.
Non-unitary states necessarily have broken time reversal symmetry.
However, note that, for example,
pairing of the form ${\bm d}_{\bm k} = (k_x+ik_y)\hat{z}$
(proposed for Sr$_2$RuO$_4$) breaks time reversal symmetry but is unitary.
The quasiparticle energies are  
\begin{equation}
E_{\bm k\pm} = \left[\varepsilon_{\bm k}^2+ \Delta_{{\bm k}\pm}^2\right]^{1/2}
\end{equation}
where
 \begin{equation}
\label{2.6}
\Delta_{{\bm k}\pm} = 
\left[|{\bm d}_{\bm k}|^2 \pm |{\bm q}_{\bm k}|\right]^{1/2}.
\end{equation}
Thus, non-unitary pairing lifts the gap degeneracy.

For triplet pairing,
the normal and anomalous quasiparticle Green's functions are\cite{Sigrist1991,correction}
\begin{equation}
\label{2.7}
\widetilde{G}({\bm k},i\omega_n)=
\frac{-[\omega_n^2+\varepsilon_{\bm k}^2+|{\bm d}_{\bm k}|^2]{\widetilde\sigma}_0+{\bm q_{\bm k}}
\cdot\widetilde{\bm\sigma}}
{[\omega_n^2+E_{{\bm k}-}^2][\omega_n^2+E_{{\bm k}+}^2]}[i\omega_n+\varepsilon_{\bm k}]
\end{equation} 
\begin{equation}
\label{2.8}
\widetilde{F}({\bm k},i\omega_n)=\frac{[\omega_n^2+\varepsilon_{\bm k}^2+|{\bm d_{\bm k}}|^2]
{\bm d_{\bm k}}-i{\bm q_{\bm k}\times{\bm d_{\bm k}}}}
{[\omega_n^2+E_{{\bm k}-}^2][\omega_n^2+E_{{\bm k}+}^2]}\cdot[i\widetilde{\bm\sigma}\widetilde{\sigma}_y]
\end{equation}
It is useful to expand these expressions as
\begin{widetext}
\begin{eqnarray}
\label{2.10}
\widetilde{G}({\bm k},\omega)=&&
\frac{\widetilde{\sigma}_0}{2}\left[\frac{u_{{\bm k}-}^2}{\omega-E_{{\bm k}-}+i\delta}+\frac{v_{{\bm k}-}^2}{\omega+E_{{\bm k}-}+i\delta}+
\frac{u_{{\bm k}+}^2}{\omega-E_{{\bm k}+}+i\delta}+\frac{v_{{\bm k}+}^2}{\omega+E_{{\bm k}+}+i\delta}\right]
\nn\\
&&-\frac{{\bm q}_{\bm k}\cdot\widetilde{\bm\sigma}}{2|{\bm q}_{\bm k}|}
\left[\frac{u_{{\bm k}-}^2}{\omega-E_{{\bm k}-}+i\delta}+\frac{v_{{\bm k}-}^2}{\omega+E_{{\bm k}-}+i\delta}-
\frac{u_{{\bm k}+}^2}{\omega-E_{{\bm k}+}+i\delta}-\frac{v_{{\bm k}+}^2}{\omega+E_{{\bm k}+}+i\delta}\right]
\end{eqnarray}
\begin{eqnarray}
\label{2.11}
\widetilde{F}({\bm k},\omega)=&&
-\frac{\widetilde{\sigma}_0}{2}\left[\frac{\widetilde{\Delta}_{\bm k}}{\Delta_{{\bm k}-}}
\left[\frac{u_{{\bm k}-}v_{{\bm k}-}}{\omega-E_{{\bm k}-}+i\delta}
-\frac{u_{{\bm k}-}v_{{\bm k}-}}{\omega+E_{{\bm k}-}+i\delta}\right]+ 
\frac{\widetilde{\Delta}_{\bm k}}{\Delta_{{\bm k}+}}
\left[\frac{u_{{\bm k}+}v_{{\bm k}+}}{\omega-E_{{\bm k}+}+i\delta}
-\frac{u_{{\bm k}+}v_{{\bm k}+}}{\omega+E_{{\bm k}+}+i\delta}\right]\right]
\nn\\
&&+\frac{{\bm q_{\bm k}\cdot\widetilde{\bm\sigma}}}{2|\bm q_{\bm k}|}\left[\frac{\widetilde{\Delta}_{\bm k}}{\Delta_{{\bm k}-}}
\left[\frac{u_{{\bm k}-}v_{{\bm k}-}}{\omega-E_{{\bm k}-}+i\delta}
-\frac{u_{{\bm k}-}v_{{\bm k}-}}{\omega+E_{{\bm k}-}+i\delta}\right]-\frac{\widetilde{\Delta}_{\bm k}}{\Delta_{{\bm k}+}}
\left[\frac{u_{{\bm k}+}v_{{\bm k}+}}{\omega-E_{{\bm k}+}+i\delta}
-\frac{u_{{\bm k}+}v_{{\bm k}+}}{\omega+E_{{\bm k}+}+i\delta}\right]\right]
\nn\\
\end{eqnarray}  
\end{widetext}
where
\begin{eqnarray}
\label{2.12}
u_{{\bm k}\pm}^2&=&\frac{1}{2}[1+\frac{\varepsilon_{\bm k}}{E_{{\bm k}\pm}}],~~v_{{\bm k}\pm}^2=\frac{1}{2}[1-\frac{\varepsilon_{\bm k}}{E_{{\bm k}\pm}}]
\nn\\
u_{{\bm k}\pm}v_{{\bm k}\pm}&=&\frac{\Delta_{{\bm k}\pm}}{2 E_{{\bm k}\pm}},~~
u_{{\bm k}\pm}^2+v_{{\bm k}\pm}^2=1
\end{eqnarray}
are the 
extended coherence factors for this particular state. Note that the following 
identity has been used in deriving the above expressions 
\begin{eqnarray}
\label{2.13}
i({\bm q}_{\bm k}\times{\bm d}_{\bm k})
\cdot\widetilde{\bm\sigma}&=&({\bm q}_{\bm k}\cdot\widetilde{\bm\sigma})
({\bm d}_{\bm k}\cdot\widetilde{\bm\sigma})-{\bm q}_{\bm k}\cdot{\bm d}_{\bm k}
\nn\\
&=&({\bm q}_{\bm k}\cdot\widetilde{\bm\sigma})({\bm d}_{\bm k}\cdot\widetilde{\bm\sigma})
\end{eqnarray}
where ${\bm q}_{\bm k}\cdot{\bm d}_{\bm k}=0$ because ${\bm q_{\bm k}}\perp{\bm d_{\bm k}}$. 
The self-energy can be included by replacing $i\omega_n$
with $i\omega_n-\Sigma(i\omega_n)$.  The retarded 
self-energy is 
$\Sigma_{ret}(\omega)=\Sigma(i\omega_n\rightarrow \omega+i\delta) =-i\Gamma(\omega)$
where the real part is assumed to be frequency independent and
absorbed in the chemical potential.  

The spectral function $\widetilde{A}^G({\bm k},\omega)$ (and similarly $\widetilde{A}^F({\bm k},\omega)$) 
is defined by
\begin{equation}
\label{2.30}
\widetilde{G}({\bm k},i\omega_n)=
\int_{-\infty}^{+\infty}d\omega \,
\frac{\widetilde{A}^G({\bm k},\omega)}{i\omega_n-\omega}.
\end{equation}
Usually, the spectral function is just $-\frac{1}{\pi}\Im{\widetilde{G}^{ret}({\bm k},\omega)}$, 
but in this case, because the Green's function has a complex
numerator, the spectral function must be extracted more carefully. Using (\ref{2.10}) and (\ref{2.11}), one finds 
\begin{widetext}
\begin{eqnarray}
\label{2.16}
\widetilde{A}^G({\bm k},\omega)=&&\frac{\Gamma(\omega)}{2\pi}\widetilde{\sigma}_0
\left[\frac{u_{{\bm k}-}^2}{(\omega-E_{{\bm k}-})^2+\Gamma^2(\omega)}+
\frac{v_{{\bm k}-}^2}{(\omega+E_{{\bm k}-})^2+\Gamma^2(\omega)}+
\frac{u_{{\bm k}+}^2}{(\omega-E_{{\bm k}+})^2+\Gamma^2(\omega)}
+\frac{v_{{\bm k}+}^2}{(\omega+E_{{\bm k}+})^2+\Gamma^2(\omega)}\right]
\nn\\
&&-\frac{\Gamma(\omega)}{2\pi}\frac{{\bm q_{\bm k}}\cdot\widetilde{\bm\sigma}}{|{\bm q_{\bm k}}|}
\left[\frac{u_{{\bm k}-}^2}{(\omega-E_{{\bm k}-})^2+\Gamma^2(\omega)}+
\frac{v_{{\bm k}-}^2}{(\omega+E_{{\bm k}-})^2+\Gamma^2(\omega)}-
\frac{u_{{\bm k}+}^2}{(\omega-E_{{\bm k}+})^2+\Gamma^2(\omega)}
-\frac{v_{{\bm k}+}^2}{(\omega+E_{{\bm k}+})^2+\Gamma^2(\omega)}\right]
\nn\\\nn\\
\end{eqnarray}
\begin{eqnarray}
\label{2.17}
& & \widetilde{A}^F({\bm k},\omega)= \nn\\
& & \frac{\Gamma(\omega)}{2\pi}\widetilde{\sigma}_0
\bigg[\frac{\widetilde{\Delta}_{\bm k}}{\Delta_{{\bm k}-}}\bigg[\frac{u_{{\bm k}-}v_{{\bm k}-}}{(\omega+E_{{\bm k}-})^2+\Gamma^2(\omega)}
-\frac{u_{{\bm k}-}v_{{\bm k}-}}{(\omega-E_{{\bm k}-})^2
+\Gamma^2(\omega)}\bigg]
+\frac{\widetilde{\Delta}_{\bm k}}{\Delta_{{\bm k}+}}\bigg[\frac{u_{{\bm k}+}v_{{\bm k}+}}
{(\omega+E_{{\bm k}+})^2+\Gamma^2(\omega)}
-\frac{u_{{\bm k}+}v_{{\bm k}+}}{(\omega-E_{{\bm k}+})^2+\Gamma^2(\omega)}\bigg]\bigg]
\nn\\
& & 
-\frac{\Gamma(\omega)}{2\pi}\frac{{\bm q_{\bm k}\cdot\widetilde{\bm\sigma}}}{|\bm q_{\bm k}|}
\bigg[\frac{\widetilde{\Delta}_{\bm k}}{\Delta_{{\bm k}-}}\bigg[\frac{u_{{\bm k}-}v_{{\bm k}-}}{(\omega+E_{{\bm k}-})^2+\Gamma^2(\omega)}
-\frac{u_{{\bm k}-}v_{{\bm k}-}}{(\omega-E_{{\bm k}-})^2+\Gamma^2(\omega)}\bigg]
+\frac{\widetilde{\Delta}_{\bm k}}{\Delta_{{\bm k}+}}\bigg[\frac{u_{{\bm k}+}v_{{\bm k}+}}
{(\omega-E_{{\bm k}+})^2+\Gamma^2(\omega)}
-\frac{u_{{\bm k}+}v_{{\bm k}+}}{(\omega+E_{{\bm k}+})^2+\Gamma^2(\omega)}\bigg]\bigg]
\nn \\
\end{eqnarray}
\end{widetext} 
with the spectral functions in hand, we can proceed to calculate the density of states and 
the transport coefficients.

\section{Density of states}
The quasiparticles density of states can be defined in terms of the spectral function as
\begin{equation}
\label{2.18}
N(\omega)=\sum_{\bm k} {\rm Tr}[\widetilde{A}^G({\bm k},\omega)]
\end{equation}
using (\ref{2.16}) we find
the general expression for the density of states in a nonunitary superconductor,
\begin{equation}
\label{2.19}
N(\omega)=\sum_{{\bm k},\pm}
\left[u_{{\bm k}\pm}^2\delta(\omega-E_{{\bm k}\pm})+v_{{\bm k}\pm}^2\delta(\omega+E_{{\bm k}\pm})\right]
\end{equation}
in the absence of impurities. 
For small $\omega$, in the vicinity of the gap node, we have
 $v_{{\bm k}\pm}\approx 0$, $u_{{\bm k}\pm}\approx 1$, and (\ref{2.19}) is
reduced to\cite{Sigrist1991} 
\begin{equation}
\label{add1}
N(\omega)\approx\sum_{{\bm k},\pm}
\delta(\omega-E_{{\bm k}\pm}).
\end{equation}
When the impurities are included the density of states becomes
\begin{equation}
\label{add2}
N(\omega)\approx\frac{\Gamma(\omega)}{\pi}\sum_{{\bm k},\pm}
\bigg[\frac{1}{(\omega-E_{{\bm k}\pm})^2
+\Gamma^2(\omega)}\bigg].
\end{equation}
It is clear from (\ref{add2}) that the residual density of states
depends on the impurity concentration through the self-energy $\Gamma(0)$.

\section{Electrical conductivity}
The DC electrical conductivity is defined by the Kubo formula \cite{Mahan2000}
\begin{equation}
\label{2.21}
{\widetilde\sigma}=-\lim_{\Omega\rightarrow 0}\frac{\Im{{\widetilde\Pi}_{ret}(\Omega)}}{\Omega}
\end{equation}
where
\begin{equation}
\label{2.22}
{\widetilde\Pi}({\bm q},i\Omega_n)=-\int_0^{\beta} d\tau e^{i\Omega_n \tau}\left<T_{\tau}{\bm j}_{\bm q}(\tau)
{\bm j}_{-\bm q}(0)\right>
\end{equation}
is the current-current correlation function. 
The electrical current is defined by 
\begin{equation}
\label{2.23}
{\bm j}({\bm q},\tau)=\frac{e}{m^{*}}\sum_{{\bm k},s}
\left[{\bm k}+\frac{\bm q}{2}\right]c_{{\bm k}+{\bm q},s}^{\dag}(\tau)\,
c_{{\bm k},s}(\tau).
\end{equation}
The current-current correlation function is therefore
\begin{eqnarray}
\label{2.28}
{\widetilde\Pi}({\bm q},i\Omega_n)=&&\frac{e^2}{m^{*2}}\sum_{{\bm k}}\left[{\bm k}+\frac{\bm q}{2}\right]^2
\nn\\
&&\frac{1}{\beta}\sum_{i\omega_n}{\rm Tr}[\widetilde{G}({\bm k},i\omega_n)
\widetilde{G}({\bm k}+{\bm q},i\omega_n+i\Omega_n)
\nn\\
&&+\widetilde{F}({\bm k},i\omega_n)
\widetilde{F}^{\dag}({\bm k}+{\bm q},i\omega_n+i\Omega_n)].
\end{eqnarray}
The conductivity 
vanishes when the self-energy is absent, 
and the contribution from the anomalous part vanishes even when the
self-energy is 
included.
In the limit 
${\bm q}\rightarrow 0$
the correlation function is
\begin{equation}
\label{2.29}
{\widetilde\Pi}(i\Omega_n)=e^2\sum_{{\bm k}}{\bm v}_F{\bm v}_F
\frac{1}{\beta}\sum_{i\omega_n}{\rm Tr}[\widetilde{G}({\bm k},i\omega_n)
\widetilde{G}({\bm k},i\omega_n+i\Omega_n)]
\end{equation}
To evaluate this correlation function we follow the approach of Refs.\
\onlinecite{Mahan2000} and \onlinecite{Durst} and rewrite 
the Green's function in terms of the spectral function
(\ref{2.16})
and sum over Matsubara frequencies.  This eventually leads to 
\begin{eqnarray}
\label{2.33}
\Im{{\widetilde\Pi}_{ret}}(\Omega)=&&-\pi e^2\sum_{{\bm k}}{\bm v}_F{\bm v}_F \int_{-\infty}^{\infty}d{\omega'}
\nn\\
&&{\rm Tr}[\widetilde{A}_{\bm k}^G(\omega')
\widetilde{A}_{\bm k}^G(\omega'+\Omega)]
\nn\\
&&[n_F(\omega')-n_F(\omega'+\Omega)].
\end{eqnarray}
Then the  DC electrical conductivity 
(\ref{2.21}) is 
\begin{eqnarray}
\label{2.34}
{\widetilde\sigma}=&&\pi e^2\sum_{{\bm k}}{\bm v}_F{\bm v}_F \int_{-\infty}^{\infty}d{\omega'}
{\rm Tr}[\widetilde{A}_{\bm k}^G(\omega')
\widetilde{A}_{\bm k}^G(\omega')]
\nn\\
&&\left[-\frac{\partial n_F(\omega')}{\partial\omega'}\right].
\end{eqnarray}
In the limit $T\rightarrow 0$ we have
$-\frac{\partial n_F(\omega')}{\partial\omega'}=\delta(\omega')$, and the 
conductivity is
\begin{eqnarray}
\label{2.35}
{\widetilde\sigma}=\pi e^2\sum_{{\bm k}}{\bm v}_F{\bm v}_F
{\rm Tr}[\widetilde{A}_{\bm k}^G(0)
\widetilde{A}_{\bm k}^G(0)].
\end{eqnarray}
Using (\ref{2.16}) we finally obtain 
the conductivity for a 
non-unitary superconductor,
\begin{equation}
\label{2.36}
{\widetilde\sigma}=\frac{e^2\Gamma_0^2}{\pi}\sum_{\bm k}{\bm v}_F{\bm v}_F
\left[\frac{1}{(\Gamma_0^2+E_{{\bm k}-}^2)^2}
+\frac{1}{(\Gamma_0^2+E_{{\bm k}+}^2)^2}\right]
\end{equation}
where $\Gamma_0=\Gamma(\omega=0)$.

\section{Thermal conductivity}
The DC thermal conductivity is defined by the Kubo formula
\cite{Durst}
\begin{equation}
\label{2.37}
\frac{{\widetilde\kappa}}{T}=-\frac{1}{T^2}\lim_{\Omega\rightarrow 0}\frac{\Im{{\widetilde\Pi}_{ret}(\Omega)}}{\Omega}.
\end{equation}
The heat current can be written in second quantization form as
\begin{eqnarray}
\label{2.38}
{\bm j}_{\bm q}(\tau)=&&-\frac{1}{2 m^{*}}\sum_{{\bm k},s}\bigg[i[{\bm k}+{\bm q}]
\frac{\partial c_{\bm k, s}^{\dag}(\tau)}{\partial\tau}c_{{\bm k}+{\bm q},s}(\tau)
\nn\\
&&-i{\bm k}\ c_{\bm k, s}^{\dag}(\tau)\frac{\partial c_{{\bm k}+{\bm q},s}(\tau)}{\partial\tau}\bigg].
\end{eqnarray}
This form is similar to (4.17) in Ref.\ [\onlinecite{Durst}] except that we 
have neglected 
the term proportional to the gap velocity, which we assume to be much
smaller than the Fermi velocity.
The current-current correlation function is then
\begin{eqnarray}
\label{2.41}
{\widetilde\Pi}(i\Omega_n)=
&&\frac{1}{\beta}\sum_{{\bm k},i\omega_n}{\bm v}_F{\bm v}_F \left[i\omega_n+\frac{i\Omega_n}{2}\right]^2
\nn\\
&&{\rm Tr}\big[\widetilde{G}_{\bm k}(i\omega_n+i\Omega_n)\widetilde{G}_{\bm k}(-i\omega_n)
\nn\\
&&-\widetilde{F}_{\bm k}(-i\omega_n)\widetilde{F}_{\bm k}^{\dag}(i\omega_n+i\Omega_n)\big]
\end{eqnarray}
As in the electrical conductivity, the anomalous part does not contribute to the
thermal conductivity.  Finally, the correlation function is expressed in terms
of the spectral function as 
\begin{eqnarray}
\label{2.43}
\Im{{\widetilde\Pi}}_{ret}(\Omega)=&&\sum_{{\bm k}}{\bm v}_F{\bm v}_F \int_{-\infty}^{\infty}d{\omega'}
\nn\\
&&{\rm Tr}[\widetilde{A}_{\bm k}^G(\omega'+\Omega)
\widetilde{A}_{\bm k}^G(-\omega')]
\nn\\
&&\left[\omega'+\frac{\Omega}{2}\right]^2\left[n_F(\omega'+\Omega)-n_F(\omega')\right].
\nn\\
\end{eqnarray}
Substituting this into the Kubo formula (\ref{2.37}) and evaluating in the limit  $\Omega\rightarrow 0$ and 
$T\rightarrow 0$, we find
\begin{equation}
\label{2.44}
\frac{{\widetilde\kappa}}{T}=\frac{\pi^2}{3}k_B^2\sum_{{\bm k}}{\bm v}_F{\bm v}_F
{\rm Tr}[\widetilde{A}_{\bm k}^G(0)
\widetilde{A}_{\bm k}^G(0)].
\end{equation}
Comparing (\ref{2.35}) and (\ref{2.44}) 
we can see that the Wiedemann-Franz law
$\frac{\kappa}{\sigma T}=\frac{\pi k_B^2}{3e^2}$ is satisfied. Explicitly, the thermal conductivity is 
\begin{equation}
\label{2.46}
\frac{{\widetilde\kappa}}{T}=\frac{k_B^2}{3}\Gamma_0^2\sum_{\bm k}{\bm v}_F{\bm v}_F
\left[\frac{1}{(\Gamma_0^2+E_{{\bm k}-}^2)^2}
+\frac{1}{(\Gamma_0^2+E_{{\bm k}+}^2)^2}\right].
\end{equation}

\section{Application to P${\rm{\bf r}}$O${\rm{\bf s_4}}$S${\rm{\bf b_{12}}}$}
As discussed in the Introduction, 
we assume that the gap function for the A phase is 
\begin{equation}
\label{Agap}
\Delta_{\bm k}=|\eta_1|\left[a^2k_y^2+b^2k_x^2\right]^{1/2},
\end{equation} 
where $a$ and $b$ are undetermined constants, 
while for the B phase it has the form 
\begin{eqnarray}
\Delta_{{\bm k}\pm}& =& \bigg[ 
\left[|\eta_1|^2 b^2 + |\eta_2|^2 a^2\right]k_x^2 
+ |\eta_1|^2 a^2 k_y^2
+|\eta_2|^2 b^2 k_z^2  \nonumber \\ 
&& \pm 2 |\eta_1||\eta_2||k_x|\sqrt{a^2b^2k_x^2+a^4k_y^2+b^4k_z^2}
\bigg]^{1/2}.\label{Bgap}
\end{eqnarray}
which is non-degenerate.\cite{Tayseer2}
The gap function in the A~phase is unitary and has two cusp point nodes in the [00$\pm$1] directions. 
The lower branch of the B phase gap function 
has four point nodes which are in the $k_y=0$ plane at the positions
$\sqrt{|\eta_1|^2b^2-|\eta_2|^2a^2}k_x = \pm |\eta_2| b k_z$ 
if $|\eta_1|^2b^2>|\eta_2|^2a^2$; else they are in the $k_z=0$ plane.
We will assume the former in our calculations.
Since we are interested 
in the very low temperature regime, we will consider only the B phase. 

The gap function of the B phase in the vicinity of nodes 
can be linearised as
\begin{equation}
\Delta_{\bm k} \approx 
v\sqrt{k_{||}^2+k_y'^2} 
\end{equation}
where $v = \sqrt{|\eta_{1}|^2b^2-|\eta_2|^2a^2}$, $k_y'=\frac{a}{b}\,k_y$ and
\begin{equation}
\label{k1}
k_{||}=\frac{\sqrt{|\eta_1|^2b^2-|\eta_2|^2a^2}}{|\eta_1|b}k_x\pm\frac{|\eta_2|a}{|\eta_1|b}k_z.
\end{equation}  
$k_{||}$ and $k_{\perp}$ (used below) are momenta 
parallel and perpendicular to the Fermi surface at the node.
The upper branch, which is degenerate with the lower branch on the line $k_x=0$ between each 
pair of nodes, is properly included with this linearisation of the gap function.
Therefore, 
we relabel the two branches of the gap function as shown in
Fig.\ 1.  Thus for any function we have
\begin{equation}
f(E_{+}) + f(E_{-}) \equiv
f(E_{1}) + f(E_{2}).
\label{branch12}
\end{equation}  
Each branch 1 and 2 has two cusp point nodes 
and the contribution to the excitation spectrum from each branch is equal. 
With this picture in mind, 
we now calculate the density of states and the transport coefficients.   

\begin{figure}[ht]
\epsfysize=1.3in
\epsfbox[90 460 530 570]{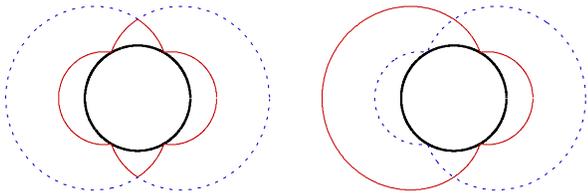}
\caption{(Color online) Gap function for the B phase of \pr\ drawn in the $k_x$-$k_z$ plane over a spherical Fermi surface (bold black).  Left: the `+' branch
is shown in blue (dashed) and the `-' branch in red (solid).  Right: 
the `1' branch is shown in blue (dashed) and the `2' branch in red (solid).}
\end{figure}

\subsection{Density of states}
The density of states  
was calculated previously in Ref.\ \onlinecite{Tayseer2} in the
absence of impurities; here we will include
the effect of impurities starting from 
(\ref{add2}).  Linearising the gap function
as 
described above,  
we find
\begin{eqnarray}
\label{2.47}
N(\omega)=\frac{\Gamma(\omega)}{\pi}2\sum_{j=1}^2\int \frac{d^3k}{(2\pi)^3}
\frac{1}{(\omega-E_{\bm k})^2+\Gamma^2(\omega)}
\end{eqnarray}    
where there is a factor of 2  because there are two branches of the gap 
function and the sum is over the two nodes in each branch.
To perform the integration we change variables to $p^2 =  
v^2(k_{||}^{2} + k_y^{'2}) + v_F^2k_{\perp}^2 \approx E_{\bm k}^2$,
\begin{equation}
\label{2.48}
N(\omega)=\frac{2\Gamma(\omega)}{\pi^3}
\frac{b}{a}\frac{1}{[|\eta_1|^2b^2-|\eta_2|^2a^2]v_F}
\int_0^{p_0}
\frac{dp\ p^2}{(\omega-p)^2+\Gamma^2(\omega)}
\end{equation}
and introduce a cutoff $p_0$.
Performing the integration we arrive at the result
\begin{eqnarray}
\label{2.49}
N(\omega)=&&\frac{2}{\pi^3}
\frac{b}{a}\frac{1}{[|\eta_1|^2b^2-|\eta_2|^2a^2]v_F}\bigg[[\omega^2-\Gamma^2(\omega)]
\nn\\
&&\bigg[\tan^{-1}\left(\frac{p_0-\omega}{\Gamma(\omega)}\right)
+\tan^{-1}\left(\frac{\omega}{\Gamma(\omega)}\right)\bigg]
\nn\\
&&+\omega\Gamma(\omega)
\ln\left(\frac{[p_0-\omega]^2+\Gamma^2(\omega)}{\omega^2+\Gamma^2(\omega)}\right)
-\omega\Gamma(\omega)
\nn\\
&&+p_0\Gamma(\omega)\bigg].
\end{eqnarray}
Setting
$\Gamma(\omega)=0$ we obtain our previous result\cite{Tayseer2} 
\begin{equation}
\label{2.51}
N(\omega)=\frac{b}{a}\frac{2\omega^2}{\pi^2v_F[|\eta_1|^2b^2-|\eta_2|^2a^2]},
\end{equation}
which has a quadratic dependence on frequency as expected for point nodes.
In the limit $\omega \rightarrow 0$ (\ref{2.49}) becomes
\begin{equation}
\label{2.52}
N(0)=\frac{2}{\pi^3}
\frac{b}{a}\frac{\Gamma_0^2}{[|\eta_1|^2b^2-|\eta_2|^2a^2]v_F}
\left[\tan^{-1}\left(-\frac{p_0}{\Gamma_0}\right)+\frac{p_0}{\Gamma_0}
\right].
\end{equation}
This is  the zero energy density of states induced by impurities.
The cut-off is normally taken to be the size of the Brillouin zone\cite{Durst} 
but it may be more physical to use the reciprocal of the
range of the single impurity potential,\cite{Balatsky1994} $p_0 \propto \lambda^{-1}$. In terms of the ratio $(p_0/\Gamma_0)$ the two limits are 
\be
\label{scat1}
 \frac{p_0}{\Gamma_0}&\gg& 1~(\rm unitary)\\
 \frac{p_0}{\Gamma_0}&\ll& 1~(\rm Born)
\ee 
In the unitary  limit the density of states is
\begin{equation}
\label{2.52a}
N(0)=\frac{2}{\pi^3}
\frac{b}{a}\frac{p_0\Gamma_0^u}{[|\eta_1|^2b^2-|\eta_2|^2a^2]v_F}
\end{equation}
where $u$ refers to unitary scattering. 
If $\Gamma_c$ is the critical 
scattering rate at which the superconductor becomes normal
at the node, then we can write (\ref{2.52a}) as
\begin{equation}
\label{2.52b}
\frac{N(0)}{N_{n}}=\frac{\Gamma_0^u}{\Gamma_c^u}=\frac{n_{imp}}{n_{imp}^c}.
\end{equation}
In the Born limit, 
the density of states
vanishes as $\Gamma_0^2$.

The presence of residual density of states, in general, gives
a contribution linear in temperature
to the specific heat and the nuclear spin relaxation rate at low
temperature. The prefactor dependence on impurity doping may be helpful in
identifying the symmetry of the order parameter.
The specific heat is\cite{Sigrist1991}
\be
\label{sh1}
C(T)=\frac{2}{T}\int_0^{\infty}d\omega\, \omega^2N(\omega)
\left[-\frac{\partial f}{\partial\omega}\right].
\ee
At low temperature this yields
\be
\label{sh2}
\frac{(C(T)/T)}{(C(T)/T)_n}
=\frac{\Gamma_0^u}{\Gamma_c^u}
\ee
and the nuclear spin relaxation rate is\cite{Sigrist1991}
\be
\label{l1}
\frac{(1/T_1)_{T}}{(1/T_1)_{n}}&=&2
\frac{T}{T_c}\int_0^{\infty}d\omega\,N(\omega)N(\omega-\omega_0)
\left[-\frac{\partial f}{\partial\omega}\right]
\nn\\
\frac{(1/TT_1)_{T}}{(1/TT_1)_{n}}&=&\frac{{\Gamma_0^u}^2}{{\Gamma_c^u}^2}.
\ee
 
\subsection{Electrical and thermal conductivities}
Beginning with  (\ref{2.36}) and making use of (\ref{branch12}),
we divide the integration into four parts,
each centred about one node in the gap function.
The factor ${\bm v}_F{\bm v}_F$ is evaluated at each node; the sum over nodes
yields
\begin{equation}
\label{2.53} 
\sum_{j=1}^4 {\bm v}_F{\bm v}_F=4 v_F^2
\left( \begin{array}{ccc}
  \frac{|\eta_2|^2a^2}{|\eta_1|^2 b^2} & 0&0 \\
 0 & 0 &0\\
0 & 0&\frac{|\eta_1|^2b^2-|\eta_2|^2 a^2}{|\eta_1|^2b^2} 
\end{array} \right)
\end{equation}
The remaining integration is the same for each part.
Performing the same change of variables as in the density of states
calculation, we find
\begin{equation}
\label{2.54}
\widetilde{\sigma}=\frac{e^2\Gamma_0^2}{2\pi^3}\frac{b}{a}\frac{\sum_{j=1}^4 {\bm v}_F{\bm v}_F}
{\left[|\eta_1|^2b^2-|\eta_2|^2a^2\right]v_F}
\int_0^{p_0}
\frac{dp\ p^2}{\left[p^2+\Gamma_0^2\right]^2}
\end{equation}
and completing the integration we get
\begin{eqnarray}
\label{2.56}
\widetilde{\sigma}&=&
\frac{e^2}{\pi^3}\ v_F\Gamma_0\left[\tan^{-1}
\left(\frac{p_0}{\Gamma_0}\right)-\frac{(p_0/\Gamma_0)}{1+(p_0/\Gamma_0)^2}\right]  \nn \\
& & \times \left( \begin{array}{ccc}
  \frac{a|\eta_2|^2}{b |\eta_1|^2 \left[b^2|\eta_1|^2-a^2|\eta_2|^2\right]} & 0&0 \\
 0 & 0 &0\\
0 & 0&\frac{1}{a b|\eta_1|^2} 
\end{array} \right)
\end{eqnarray}
This is the impurity induced  DC electrical conductivity 
for the B phase of \pr.  The thermal 
conductivity can be easily obtained by using
the Wiedemann-Franz law. In the unitary limit ($\frac{p_0}{\Gamma_0}\gg 1$), 
the term which includes 
$\tan^{{-1}}\left(\frac{p_0}{{\Gamma_0}}\right)=\frac{\pi}{2}$ will dominate, 
the conductivities become
\begin{equation}
\label{2.57}
\widetilde{\sigma}=\frac{e^2}{2\pi^2}\ v_F{\Gamma_0^u} 
\left( \begin{array}{ccc}
  \frac{a|\eta_2|^2}{b |\eta_1|^2 \left[b^2|\eta_1|^2-a^2|\eta_2|^2\right]} & 0&0 \\
 0 & 0 &0\\
0 & 0&\frac{1}{a b|\eta_1|^2} 
\end{array} \right)
\end{equation}
and
\begin{equation}
\label{2.58}
\frac{\widetilde{\kappa}}{T}=\frac{k_B^2}{6\pi}\ v_F\Gamma_0^u 
\left( \begin{array}{ccc}
  \frac{a|\eta_2|^2}{b |\eta_1|^2 \left[b^2|\eta_1|^2-a^2|\eta_2|^2\right]} & 0&0 \\
 0 & 0 &0\\
0 & 0&\frac{1}{a b|\eta_1|^2} 
\end{array} \right).
\end{equation}
         
Thus the conductivities in the B phase of \pr\  are non-universal 
(dependent on impurity concentration) 
for unitary scattering but vanish in the Born limit.
The conductivity tensor has two  inequivalent
components, $\sigma_{xx}$ and $\sigma_{zz}$ due to the off-axis nodal positions
{\em and} the choice of a
particular domain of superconducting phase.   This 
domain is represented by order parameter components
$(0,i|\eta_2|,|\eta_1|)$.  If all six domains are present then
all diagonal components of the conductivity tensor will be equal.
The $\sigma_{xx}$ component is proportional to the 
parameter $|\eta_2|$ which is absent in the unitary A phase. 
Therefore, measurement of residual conductivities in a domain-pinned
set-up, such as the one used in directional dependent thermal conductivity
measurements\cite{Izawa2003} could determine the direction of
nodes.  Of all the possible SC states in tetrahedral systems, $D_2(E)$, with
OP components $(0,i|\eta_2|,|\eta_1|)$, is the only one with off-axis nodes.\cite{Sergienko2004,Tayseer2}

\subsection{Discussion}

There have been several studies on Ru and La doped samples,\cite{Frederick2004, Nishiyama2005,Yogi2006,Rotundu2006,Maple2006} with the surprising result
that Ru substitution  leads to a doping-dependent
residual density of states and resistivity,\cite{Frederick2004, Nishiyama2005} while La substitution does not.\cite{Yogi2006}
In \pr,  it is speculated that quadrupolar fluctuations
 of the Pr ions play a  role  similar to the magnetic fluctuations of Ce and U
ions in other heavy fermion superconductors, thus substitution of the Pr ions
by La would be expected to produce unitary scatterers.
However, in contrast to Eq.\ \ref{l1}, there is no dependence on doping
on NQR relaxation rate beyond the La concentration $x=0.05$.

Both Pr$_{1-x}$La$_x$Os$_4$Sb$_{12}$ and Pr(Os$_{1-x}$Ru$_x$)$_4$Sb$_{12}$
are superconducting for the entire range of $x$, and both become 
$s$-wave superconductors at some intermediate value of $x$.
In the Ru doped series, $T_c$ has a minimum at $x=0.6$, with a 
leveling off of the specific heat at the same value.  This suggests that
a phase transition between triplet and singlet superconductivity occurs 
at $x \approx 0.6$, with possibly a region of co-existence of these two
phases.\cite{Sergienko2004b}  A 0.4\% change in lattice constant occurs
between \pr\ and PrRu$_4$Sb$_{12}$,\cite{Frederick2004} and effects due
to quadrupolar fluctuations appear to be absent in PrRu$_4$Sb$_{12}$.
In the La doped
series, $T_c$ decreases linearly along the entire range of $x$, while
the specific heat levels off at $x\approx 0.3$.

According to (\ref{2.52a}) and (\ref{2.57}), the dependence of the 
residual density of states and 
resistivity on Ru doping suggests that  the  scattering from Ru
ions is unitary.
Unitary scattering due to the substitution of Os by Ru may be explained 
by noting that quadrupolar fluctuations of the Pr ions 
are charge density fluctuations and 
will couple to, and possibly be enhanced by, quadrupolar lattice vibration
modes.  The change in lattice constant that accompanies Ru doping 
will alter the 
quadrupole moment of those modes.  In addition, Ru substitution has a strong
effect on the low-lying crystal electric field (CEF) levels of the Pr ions
which eventually removes quadrupole fluctuations.\cite{Maple2006}
La substitution produces a much smaller change in lattice constant and 
has a much weaker effect on the Pr CEF levels.   Nevertheless,
it is still difficult to explain why there is
no dependence at all on the  La concentration.

\section{Summary and Conclusions}
It is evident from (\ref{add2}), (\ref{2.36}) and (\ref{2.46}) that the main
effect of a  non-unitary superconducting state is a lifting of the
gap degeneracy, and that this would be observed as multi-gap behaviour 
similar to what could be expected for multi-band superconductivity.
There are, however, some differences which we outline here. 
We base the following discussion on the unitary state $D_2(C_2)\times{\cal K}$ and the non-unitary state
$D_2(E)$, with order parameter components $(0,0,|\eta_1|)$ and 
$(0,i|\eta_2|,|\eta_1|)$ respectively.  There
are many other states, but all the rest are either nodeless,
or else they have a $C_3$ symmetry element which has been positively ruled out
by experiment.\cite{Huxley2004} 

In a multi-band superconductor with a single $T_c$ 
the symmetry of the superconducting
order parameter should either be the same on  both bands, or possibly,
superconductivity on one band is a secondary order parameter to 
superconductivity on the other.  The alternative, which is
the simultaneous appearance of two different order parameters, 
would be unprecedented.  This means that the symmetries of superconducting
states on the different bands
should either be the same, or have a group-subgroup relation.  
For example, in MgB$_2$, the archetypal multi-band superconductor, 
s-wave superconductivity is observed as a full gap for
both bands.
The best candidates for nodal superconductivity in the 
triplet channel in \pr\ are the unitary state $D_2(C_2)\times{\cal K}$ 
and  the non-unitary state $D_2(E)$, and
neither of these has secondary order parameters.\cite{Sergienko2004}
Therefore multi-band superconductivity entails nodes at the same
places for both gaps, unless that part of the Fermi surface is missing.
On the other hand, the non-unitary superconducting state has nodes in 
the lower branch and a fully gapped upper branch.
This difference may help to distinguish these two possibilities.

To summarise, we have found general expressions for the 
residual density of states and electrical and thermal 
conductivities due to impurity scattering, and we have applied the
results to the non-unitary B phase of \pr.  The nodal positions of the non-unitary state $D_2(E)$
are unique among all the superconducting states for
crystals with tetrahedral symmetry,\cite{Sergienko2004,Tayseer2}
in that they are not found on a symmetry axis.
Inequivalent diagonal components of the conductivity 
tensor would be an unmistakable signature of such a state.

\begin{acknowledgments}
This work was supported by NSERC of Canada.
\end{acknowledgments}

\appendix



\end{document}